\documentclass[a4paper,12pt]{article}
\usepackage{amssymb, amsmath}
\usepackage[curve]{xypic}

\providecommand{\Hom}{\textnormal{Hom}}

\providecommand{\Hol}{\textnormal{Hol}}

\begin{document}

\begin{titlepage}
\titlepage
$ $\vskip 0.5cm
\centerline{ \bf \LARGE Topics on the geometry of D-brane}
\vskip 0.7cm
\centerline{ \bf \LARGE charges and Ramond-Ramond fields}
\vskip 1cm
\centerline{ \bf \Large Part II: Low-degree Ramond-Ramond fields}
\vskip 1.7truecm

\begin{center}
{\bf \large Fabio Ferrari Ruffino}
\vskip 1.5cm
\em 
ICMC - Universidade de S\~ao Paulo \\ 
Avenida Trabalhador s\~ao-carlense, 400 \\
13566-590 - S\~ao Carlos - SP, Brasil

\vskip 2cm

\large \bf Abstract
\end{center}

We discuss the geometry of the Ramond-Ramond fields of low degree via the language of abelian $p$-gerbes, working in type II superstring theory with vanishing $H$-flux. A D$p$-brane is a source for the Ramond-Ramond field strength $G_{p+2}$, which can be thought of as the curvature of a connection on a $p$-gerbe with $U(1)$-band. The highest degree forms of the connection are the local Ramond-Ramond potentials $C_{p+1}$. This picture is clear for $p \geq 0$, a $0$-gerbe being a line bundle. Actually, one can encounter also lower degree Ramond-Ramond field strengths: $G_{1}$ in IIB superstring theory, and $G_{0}$ in IIA superstring theory. Although the geometrical nature of these objects is simple in itself, it is interesting to explicitly clarify the analogy with the geometry of the higher degree fields and potentials, introducing the notions of $(-1)$-gerbes with connection and $(-2)$-gerbes. We define these gerbes and we show that they are better described by a variant of ordinary $\check{\rm{C}}$ech cohomology.

\vskip2cm

\vskip1.5\baselineskip

\vfill
 \hrule width 5.cm
\vskip 2.mm
{\small 
\noindent }
\begin{flushleft}
ferrariruffino@gmail.com
\end{flushleft}
\end{titlepage}

\newtheorem{Theorem}{Theorem}[section]
\newtheorem{Lemma}[Theorem]{Lemma}
\newtheorem{Corollary}[Theorem]{Corollary}
\newtheorem{Rmk}[Theorem]{Remark}
\newtheorem{Def}{Definition}[section]
\newtheorem{ThmDef}[Theorem]{Theorem - Defintion}

\tableofcontents

\section{Introduction}

In the paper \cite{FR} we discussed some topics on the geometry of D-brane charges and Ramond-Ramond fields. We used homology and cohomology in analogy with the theory of electromagnetism, since this is the most natural approach to this topic, before introducing more refined instruments as K-theory. We have shown that, for $p \geq 0$, the Ramond-Ramond potentials $C_{p+1}$ are the highest degree components of a connection on a $p$-gerbe with $U(1)$-band (considering line bundles as $0$-gerbes), while the associated field-strength $G_{p+2}$ is the corresponding curvature, satisfying the Dirac quantization condition.

The geometry of abelian $p$-gerbes with connection can be described in a unitary way using Deligne cohomology. A $p$-gerbe is topologically classified by its first Chern class, and a connection is completely determined by its holonomy. The latter is a function from the space of $(p+1)$-submanifolds without boundary to $U(1)$, and a section of a line bundle over the space of $(p+1)$-submanifolds with boundary: physically it corresponds to the Wess-Zumino action, i.e.\ to the the minimal coupling of the local Ramond-Ramond potentials to the D-brane world-volume.

The geometry of these objects is clear for $p \geq 0$, considering line bundles as $0$-gerbes; in these cases the Ramond-Ramond field strength $G_{p+2}$ has degree at least 2. Actually, one can encounter lower degree Ramond-Ramond fields: in IIB superstring theory there is also the 1-form field-strength $G_{1}$, while in IIA superstring theory there is a $0$-form field strength $G_{0}$, which corresponds in the semiclassical limit to the $0$-form field-strength of the massive (or Romans) IIA supergravity \cite{Polchinski}. A $0$-form is a function, and actually $G_{0}$ turns out to be constant and integral. For $G_{1}$ the local potentials are the functions $C_{0}$, while $G_{0}$ does not have any potential. The geometry of these objects is simple in itself, but it is interesting to explicitly show the analogy with the higher degree cases, in order to have a unitary picture for the Ramond-Ramond fields. We thus introduce the notion of $(-1)$-gerbes and $(-2)$-gerbes, and we show how to deal with connections, holonomies, curvatures and first Chern class for such gerbes. We show that in order to include also these low-degree gerbes it is better to introduce a variant of the ordinary $\check{\rm{C}}$ech cohomology, which we call hat-cohomology.

The paper is organized as follows. In section \ref{RRpGerbes} we briefly recall how to describe the Ramond-Ramond fields using the language of $p$-gerbes with connection. In section \ref{HatCohomology} we show how to describe the low-degree Ramond-Ramond fields, introducing a variant of $\check{\rm{C}}$ech cohomology with is suitable to describe $p$-gerbes with $U(1)$-band for any $p \in \mathbb{Z}$. In section \ref{DiffGeometry} we include the cases $p = -1$ and $p = -2$ in the theory of connections, holonomies and curvatures on $p$-gerbes, showing some simple examples. In section \ref{Conclusions} we draw our conclusions.

\section{Ramond-Ramond fields and $p$-gerbes}\label{RRpGerbes}

If we consider the classical magnetic monopole in 3+1 space-time dimensions \cite{Naber}, it is well-known that, because of the Dirac quantization condition, the field strength $F_{\mu\nu}$ can be considered as the curvature of a connection on a gauge bundle on $\mathbb{R}^{3} \setminus \{0\}$, whose first Chern class, belonging to $H^{2}(\mathbb{R}^{3} \setminus \{0\}, \mathbb{Z}) \simeq \mathbb{Z}$, corresponds to the magnetic charge fixed in the origin. If we argue in the same way for a monopole in a generic space-time dimension $n+1$, we need a gauge invariant integral $(n-1)$-form $F_{\mu_{1}\ldots\mu_{n-1}}$, whose integral on an $(n-1)$-dimensional sphere around the origin of $\mathbb{R}^{n}$ is the magnetic charge (up to a normalization constant). Hence, because of the Dirac quantization condition, such a field strength can be thought of as the curvature of a connection on a $(n-3)$-gerbe with $U(1)$-band, whose first Chern class, belonging to $H^{n-1}(\mathbb{R}^{n} \setminus \{0\}, \mathbb{Z}) \simeq \mathbb{Z}$, corresponds to the charge fixed in the origin. That's why $p$-gerbes naturally arise when dealing with monopoles in a space-time of generic dimension. Since a D-brane, at a semiclassical level, can be thought of as a generalized magnetic monopole, it follows that the Ramond-Ramond potentials $C_{\mu_{1}\ldots\mu_{p+1}}$ and field strengths $G_{\mu_{1}\ldots\mu_{p+2}}$ can be thought of respectively as a connection and its curvature on an abelian $p$-gerbe.

\subsection{Gerbes and Cech hypercohomology}\label{HyperC}

We refer to \cite{Hitchin} for an introduction to gerbes, and to \cite{Brylinski} for a more formal discussion. The main features of an abelian $p$-gerbe on a manifold $X$ are:
\begin{itemize}
	\item the equivalence class (up to isomorphism) of the $p$-gerbe $[\mathcal{G}] \in \check{H}^{p+1}(X, \underline{U}(1))$, for $\underline{U}(1)$ the sheaf of $U(1)$-valued smooth functions;
	\item the first Chern class $c_{1}(\mathcal{G}) \in H^{p+2}(X, \mathbb{Z})$, which completely determines the equivalence class of the $p$-gerbe;
	\item the equivalence class of the couple gerbe+connection $[(\mathcal{G}, \nabla)] \in \check{H}^{p+1}(X, \underline{U}(1) \overset{\tilde{d}}\longrightarrow \Omega^{1}_{\mathbb{R}} \overset{d}\longrightarrow \cdots \overset{d}\longrightarrow \Omega^{p+1}_{\mathbb{R}})$ (we will recall in the following the definition of this group);
	\item the holonomy of the connection $\Hol_{\nabla}: L^{p+1}X \rightarrow U(1)$, for $L^{p+1}X$ the space of $(p+1)$-dimensional submanifolds of $X$ without boundary;
	\item the curvature of the connection, which is a closed integral $(p+2)$-form $F \in \Omega^{p+2}_{\mathbb{R}}(X)$ such that $[F]_{dR} \simeq c_{1}(\mathcal{G}) \otimes_{\mathbb{Z}} \mathbb{R}$;
	\item flat connections are classified by $H^{p+1}(X, U(1))$, for $U(1)$ the \emph{constant} sheaf, and the first Chern classes of the $p$-gerbes on which they can be defined belong to the torsion subgroup of $H^{p+2}(X, \mathbb{Z})$.
\end{itemize}
Let us briefly comment these features. Since the equivalence class belongs to the $(p+1)$-cohomology group, with respect to a good cover $\mathfrak{U} = \{U_{\alpha}\}_{\alpha \in I}$ of $X$ the transition functions are defined on the $(p+2)$-intersections, thus they are functions $g_{\alpha_{0} \ldots \alpha_{p+1}}: U_{\alpha_{0}} \cap \ldots \cap U_{\alpha_{p+1}} \rightarrow U(1)$. To compute the first Chern class, we use the Bockstein map of the sequence $0 \rightarrow \mathbb{Z} \rightarrow \underline{\mathbb{R}} \rightarrow \underline{U}(1) \rightarrow 1$ in degree $p+1$. Concretely, we write the transition functions as $g_{\alpha_{0} \cdots \alpha_{p+1}} = e^{2\pi i \rho_{\alpha_{0} \cdots \alpha_{p+1}}}$, so that $\check{\delta}\{\rho_{\alpha_{0} \cdots \alpha_{p+1}}\} = \{c_{\alpha_{0} \cdots \alpha_{p+2}}\}$ with $c_{\alpha_{0} \cdots \alpha_{p+2}} \in \mathbb{Z}$, and we get $c_{1}(\mathcal{G}) = [\{c_{\alpha_{0} \cdots \alpha_{p+2}}\}] \in \check{H}^{p+2}(X, \mathbb{Z})$. A $p$-gerbe with connection is an element of the $\rm\check{C}$ech hypercohomology group (called Deligne cohomology group):
\begin{equation}\label{HyperCGroup}
	\check{H}^{p+1}(X, \underline{U}(1) \overset{\tilde{d}}\longrightarrow \Omega^{1}_{\mathbb{R}} \overset{d}\longrightarrow \cdots \overset{d}\longrightarrow \Omega^{p+1}_{\mathbb{R}})
\end{equation}
for $\tilde{d} = \frac{1}{2\pi i} d \circ \log$. To define this group, we consider the $\rm\check{C}$ech double complex with respect to a good cover $\mathfrak{U}$ associated to the complex $\underline{U}(1) \rightarrow \Omega^{1}_{\mathbb{R}} \rightarrow \cdots \rightarrow \Omega^{p+1}_{\mathbb{R}}$, and the group \eqref{HyperCGroup} is the $(p+1)$-th cohomology group of the associated total complex. Hence, the group of $(p+1)$-cochains is $\check{C}^{p+1}(\mathfrak{U}, \underline{U}(1) \rightarrow \Omega^{1}_{\mathbb{R}} \rightarrow \cdots \rightarrow \Omega^{p+1}_{\mathbb{R}}) = \check{C}^{p+1}(\mathfrak{U}, \underline{U}(1)) \oplus \check{C}^{p}(\mathfrak{U}, \Omega^{1}_{\mathbb{R}}) \oplus \cdots \oplus \check{C}^{0}(\mathfrak{U}, \Omega^{p+1}_{\mathbb{R}})$, and a representative hypercocycle of a gerbe with connection is a sequence $(g_{\alpha_{0} \cdots \alpha_{p+1}}, (C^{(p)}_{1})_{\alpha_{0} \cdots \alpha_{p}}, \ldots, (C'_{p})_{\alpha_{0}\alpha_{1}}, (C_{p+1})_{\alpha_{0}})$, satisfying the conditions:
\begin{equation}\label{PGerbesCocycle}
\begin{array}{l}
	(C_{p+1})_{\beta} - (C_{p+1})_{\alpha} = (-1)^{p+1} d(C'_{p})_{\alpha\beta}\\
	(C'_{p})_{\alpha\beta} + (C'_{p})_{\beta\gamma} + (C'_{p})_{\gamma\alpha} = (-1)^{p} \, d(C''_{p-1})_{\alpha\beta\gamma}\\
	\ldots\\
	\check{\delta}^{p}(C^{(p)}_{1})_{\alpha_{0}\ldots \alpha_{p}} = \frac{1}{2\pi i} d\log g_{\alpha_{0}\ldots \alpha_{p+1}}\\
	\check{\delta}^{p+1}g_{\alpha_{0}\ldots \alpha_{p+1}} = 1.
\end{array}
\end{equation}
The local forms $dC_{p+1}$ glue to a gauge-invariant one $G_{p+2}$, which is the curvature of the $p$-gerbe, but the data of the superstring background must include a complete equivalence class represented by $(g_{\alpha_{0} \cdots \alpha_{p+1}}, (C^{(p)}_{1})_{\alpha_{0} \cdots \alpha_{p}}, \ldots, (C'_{p})_{\alpha_{0}\alpha_{1}}, (C_{p+1})_{\alpha_{0}})$, not only the top-forms $C_{p+1}$. As for line bundles, the correspondence $[G_{p+2}]_{dR} \simeq c_{1}(\mathcal{G}) \otimes_{\mathbb{Z}} \mathbb{R}$ holds, in particular the Dirac quantization condition applies for any $p$. From a physical point of view, Deligne cohomology describes gauge transformations. Conditions \eqref{PGerbesCocycle} specify how the local potentials glue on the intersections, and this concerns a single representative $(g_{\alpha_{0} \cdots \alpha_{p+1}}, (C^{(p)}_{1})_{\alpha_{0} \cdots \alpha_{p}}, \ldots, (C_{p+1})_{\alpha_{0}})$ of the equivalence class. There are also possible gauge transformations consisting in the addition of a coboundary, the latter being of the form:
	\[\begin{split}
	\delta\{&h_{\alpha_{0} \ldots \alpha_{p}}, (C^{(p-1)}_{1})_{\alpha_{0} \ldots \alpha_{p-1}}, \ldots, (C_{p})_{\alpha_{0}}\} = \\
	&\{\check{\delta}^{p}h_{\alpha_{0} \ldots \alpha_{p}}, \tilde{d}h_{\alpha_{0} \ldots \alpha_{p}} + \check{\delta}^{p-1}(C^{(p-1)}_{1})_{\alpha_{0} \ldots \alpha_{p-1}}, \ldots, (-1)^{p-1}d(C^{(1)}_{p-1})_{\alpha_{0}\alpha_{1}} - (C_{p})_{\alpha_{0}} + (C_{p})_{\alpha_{1}} \}.
\end{split}\]
There real data is not the single representative, but the cohomology class itself, since it is determined by the two real physical observables: the field strength (corresponding to the field $F$ in electromagnetism) and the holonomy of the connection or Wess-Zumino action (corresponding in electromagnetism to the phase difference measured in the context of the Aranhov-Bohm effect).

We can define the holonomy (i.e.\ the exponential of the Wilson loop) for $p$-gerbes generalizing the definition for line bundles. For $p = 1$ we refer to \cite{BFS}, for $p > 1$ it becomes more complicated to write down, but the properties are the same: it is a well-defined function on the space of $(p+1)$-submanifolds without boundary, while it is the section of a line bundle over the space of $(p+1)$-submanifolds with boundary. To obtain a geometrical intuition for such objects one must deal with sheaves of $p$-categories (for $p = 1$ a good reference is \cite{Brylinski}), but the language of Deligne cohomology is enough for our purposes.

\subsection{Wess-Zumino action and D-branes duality}

We work in type II superstring theory on a 10-dimensional background $X$, with vanishing $H$-flux. The Ramond-Ramond field strength $G_{p+2}$ can be thought of as the curvature of a connection on an abelian $p$-gerbe, while the local Ramond-Ramond potentials $C_{p+1}$ are the highest degree components of the connection. There is a natural duality between D$p$-branes and D$(6-p)$-branes, because a D$(6-p)$-brane minimally couples, via the Wess-Zumino action, to the potential of the Ramond-Ramond field whose source is a D$p$-brane.\footnote{In the democratic formulation of supergravity, D$p$-branes and D$(6-p)$-branes are also sources of Hodge-dual Ramond-Ramond field-strengths $G_{8-p}$ and $G_{p+2}$.} In fact, let us consider a D$p$-brane with $(p+1)$-dimensional world-volume $WY_{p}$, for $0 \leq p \leq 6$. The violated Bianchi identity is:
\begin{equation}\label{MaxwellDBrane}
	dG_{8-p} = q_{p} \cdot \delta(WY_{p}) \qquad dG_{p+2} = 0
\end{equation}
where $q_{p}$ is the charge, or equivalently, the number of D-branes in the stack. To compute the charge from the background data, we consider a linking manifold\footnote{A linking manifold is the boundary of a manifold intersecting $WY_{p}$ transversally in isolated points of its interior.} $L$ of $WY_{p}$ in $S$ with linking number $l$, so that, from \eqref{MaxwellDBrane}, we get:
\begin{equation}\label{ChargeQp}
	q_{p} = \frac{1}{l} \int_{L} G_{8-p}.
\end{equation}
Thus the D$p$-brane is a charged source for the Ramond-Ramond field-strength $G_{8-p}$. Let us now consider a D$(6-p)$-brane with $(7-p)$-dimensional world-volume $WY_{6-p}$, moving in the $G_{8-p}$-field generated by the D$p$-brane. The Wess-Zumino action is:
\begin{equation}\label{WessZumino}
	S_{WZ} = q_{6-p} \int_{WY_{6-p}} C_{7-p}
\end{equation}
where the integral (actually, its exponential) should be interpreted as the holonomy of the $(6-p)$-gerbe connection on $WY_{6-p}$ \cite{FR}. Since $dC_{7-p} = G_{8-p}$, the D$(6-p)$-brane minimally couples to the connection associated to the field generated by the D$p$-brane.

This duality between D$p$-branes and D$(6-p)$-branes is clear for $0 \leq p \leq 6$. Let us now consider $p \geq 7$. In this case, we need to consider D-branes of negative dimension \cite{Polchinski}. For a D$7$-brane, the charge can be computed from \eqref{ChargeQp} with $p = 7$, so that the generated field is $G_{1}$. Its potential is $C_{0}$, a locally-defined 0-form, thus a locally-defined function. The dual D$(-1)$-brane has 0-dimensional world-volume, thus it is a D-instanton, whose action is by analogy with \eqref{WessZumino}:
	\[S_{WZ} = q_{-1} \cdot C_{0}(WY_{-1}),
\]
where $WY_{-1}$ is a point in which we evaluate $C_{0}$. What is the analogy between $C_{0}$, which is a locally-defined function, and the other potentials, which are connections on a $p$-gerbe? Moreover, the connections on a $p$-gerbe are determined, up to gauge transformations, by their holonomy: what is the holonomy of $C_{0}$? We will try to clarify this in the following chapters.

For a D$8$-brane, the charge can be computed from \eqref{ChargeQp} with $p = 8$, so that the generated field is $G_{0}$. Since it is already a function, it has no potential. As all the other field strengths it must be closed, hence constant. The Dirac quantization condition is usually interpreted as the requirement that $G_{0}$ is an integer number, even if this not the most general case, as we show in the following. The dual D$(-2)$-brane has $(-1)$-dimensional world-volume, thus it does not exist. This is coherent with the fact that there is no potential. In other words, there are no branes minimally coupling to the field generated by a D$8$-brane: in fact, a D$8$-brane has a $9$-dimensional world-volume which breaks in two pieces the space-time, and the field generated is the locally constant field $G_{0}$ whose variation across the world-volume is the charge; each of the two parts is independent, thus it is impossible to find moving charges measuring such a difference since they cannot cross the wall.

Finally, we should consider D$9$-branes, but, as is well-known, they have no charge: that's because, since they fill the whole space-time, there is no possibility for the fluxes to go at infinity. It is the analogue of an electron on a 2-sphere: we must put an anti-electron in another point, otherwise the theory is inconsistent. Actually, the Wess-Zumino action should be \eqref{WessZumino} with $p = -3$, but $C_{10}$ is a volume form which is non-dynamical and, as we have already said, $q_{9}$ has no meaning.

\paragraph{}Summarizing:
\begin{itemize}
	\item for $0 \leq p \leq 6$, the Ramond-Ramond field $G_{8-p}$ generated by a D$p$-brane is a curvature of a connection on a $p$-gerbe, whose holonomy is the Wess-Zumino action of a test D$(6-p)$-brane;
	\item for $p = 7$, the Ramond-Ramond field $G_{1}$ generated by a D$7$-branes is a 1-form, whose potential is a locally defined function $C_{0}$; the Wess-Zumino action of a D$(-1)$-brane, or D-instanton, is the value of $C_{0}$ on the world-volume, which is a point;
	\item for $p = 8$, the Ramond-Ramond field $G_{0}$ generated by a D$8$-branes is a constant integral function, without potential;
	\item for $p \geq 9$ there are no D-branes.
\end{itemize}
Thus, we should find a unitary description of abelian $p$-gerbes with connection (where $0$-gerbes are line bundles), leading for $p = -1$ to $1$-forms with potential, for $p = -2$ to a constant function without potential, and for $p \leq -3$ just to the trivial case. Moreover, we should extend in a unitary way the definition of holonomy and first Chern class to these cases.

\subsection{The low degree Ramond-Ramond fields}\label{LowDegRR}

\subsubsection{The Ramond-Ramond potential $C_{0}$}

We now consider in detail the case of $C_{0}$, that should correspond to the case of $(-1)$-gerbes: we start with the topological description, while we will discuss later how to define connections.

By analogy, it is natural to define a $(-1)$-gerbe (up to isomorphism) as a cohomology class in $\check{H}^{0}(X, \underline{U}(1))$, i.e.\ \emph{as a function} $f: X \rightarrow U(1)$. We can define the first Chern class of such a function as a class in $\check{H}^{1}(X, \mathbb{Z})$, computed via the Bockstein map in degree $0$ of the sequence:
\begin{equation}\label{ExSeqZRS1}
	0 \longrightarrow \mathbb{Z} \longrightarrow \underline{\mathbb{R}} \longrightarrow \underline{U}(1) \longrightarrow 1.
\end{equation}
We concretely compute it as usual: we define $f\vert_{U_{\alpha}} = e^{2\pi i \rho_{\alpha}}$, so that $\rho_{\alpha} - \rho_{\beta} = \rho_{\alpha\beta} \in \mathbb{Z}$, and $[\{\rho_{\alpha\beta}\}] \in \check{H}^{1}(X, \mathbb{Z})$ is $c_{1}(f)$. In string theory the local logarithms $\rho_{\alpha}$ are the Ramond-Ramond potentials $(C_{0})_{\alpha}$, while the globally defined exponential $f = e^{2\pi i (C_{0})_{\alpha}}$, which we will call $\tilde{C}_{0}$, is the path-integral measure corresponding to the Wess-Zumino action of a D$(-1)$-brane. Can we give a more geometrical characterization of the Chern class of a function with values in $U(1)$? Since $H^{1}(U(1), \mathbb{Z}) \simeq \mathbb{Z}$, given a function $f: X \rightarrow U(1)$ we can consider the pull-back of a generator $f^{*}(\pm 1) \in H^{1}(X, \mathbb{Z})$. This turns out to be the first Chern class of $f$ \cite{Hitchin}, once that we have fixed an orientation of $U(1)$ in order to distinguish $1$ and $-1$:
\begin{Lemma}\label{c1f} Let us give to $U(1) \subset \mathbb{C}$ the counter-clockwise orientation, so that we fix an oriented generator $1 \in H^{1}(U(1), \mathbb{Z}) \simeq \mathbb{Z}$. Then, for $f: X \rightarrow U(1)$:
	\[c_{1}(f) = f^{*}(1) \; .
\]
\end{Lemma}
The proof is in the appendix \ref{ChernClassF}. We have thus seen that it seems natural to define $(-1)$-gerbes on $X$ as functions $f: X \rightarrow U(1)$, and that the first Chern class, computed in analogy with higher degree gerbes, has a clear geometric interpretation. The problem is that, since $\check{H}^{0}(X, \underline{\mathbb{R}})$ is non-zero, while $\check{H}^{p}(X, \underline{\mathbb{R}}) = 0$ for any $p \geq 1$ (because of the existence of partitions of unity), the Bockstein map $\beta_{0}$ associated to the sequence \eqref{ExSeqZRS1} is not injective, although it is still surjective. This means that the first Chern class of a $(-1)$-gerbe does not characterize the $(-1)$-gerbe up to isomorphism, even topologically. We see from the exactness of \eqref{ExSeqZRS1} that the kernel of $\beta_{0}$ is made by functions which admits a global logarithm, i.e.\ functions such that $f = e^{2\pi i \rho}$ for some $\rho: X \rightarrow \mathbb{R}$. Of course if we compute the Chern class as above we obtain 0, since $\rho_{\alpha} = \rho_{\beta}$ for every $U_{\alpha\beta}$. We can give a topological characterization of such functions, thanks to the following simple lemma.
\begin{Lemma}\label{LogHomotopy} A function $f: X \rightarrow U(1)$ admits a global logarithm if and only if it is homotopic to the constant map $1$.
\end{Lemma}
\textbf{Proof:} If $f = e^{2 \pi i \rho}$, since $\mathbb{R}$ is contractible we can find a homotopy $F$ between $\rho$ and the zero map: for example, we can link $\rho(x)$ to $0$ with the segment joining them. Then, $e^{2\pi i F}$ is a homotopy between $f$ and $1$. Viceversa, the result follows from the homotopy lifting properties of coverings \cite{Hatcher}, but we can also prove it from lemma \ref{c1f}. In fact, since the pull-back in cohomology is homotopy-invariant, if $f$ is homotopic to $1$ then $c_{1}(f) = 1^{*}(1) = 0$, thus, from the exactness of \eqref{ExSeqZRS1}, we deduce that it admits a global logarithm. $\square$

\paragraph{}It follows that $c_{1}(f) = c_{1}(g)$ if and only if $f$ is homotopic to $g$, hence \emph{the first Chern class of a function uniquely characterizes its equivalence class up to homotopy}. Therefore, if we want to keep the analogy between $(-1)$-gerbes and $p$-gerbes for $p \geq 0$, we are in the following situation:
\begin{itemize}
	\item if we want to keep the classification of $p$-gerbes via classes in $\check{H}^{p+1}(X, \underline{U}(1))$, we must define a $(-1)$-gerbe, even up to isomorphism, as a function $f: X \rightarrow U(1)$;
	\item if we want to preserve the topological classification via the first Chern class in $\check{H}^{p+2}(X, \mathbb{Z})$, we must define a $(-1)$-gerbe, up to isomorphism, as a homotopy class of functions $[f: X \rightarrow U(1)]$.
\end{itemize}
In the following subsection we see that there is a good reason to prefer the second case, which will be confirmed studying connections and holonomies.

\subsubsection{The Ramond-Ramond field strength $G_{0}$}

The Ramond-Ramond field strength $G_{0}$, in IIA superstring theory, is a 0-form without potential. Thus, we must also consider $(-2)$-gerbes, although they are very simple. Hence:
\begin{itemize}
	\item if we keep the first analogy with $p$-gerbes, we must define a $(-2)$-gerbe as a class in $\check{H}^{-1}(X, \underline{U}(1))$, which is $0$, so we would not be able to define non-trivial $(-2)$-gerbes;
	\item if we keep the second analogy with $p$-gerbes, we must define a $(-2)$-gerbe as an object classified up to isomorphism by $\check{H}^{0}(X, \mathbb{Z})$, the latter being isomorphic to $\mathbb{Z}$ for $X$ connected: this is correct, since $G_{0}$ is a pure quantized curvature, expressed as a $\mathbb{Z}$-valued function, necessarily constant since $\mathbb{Z}$ is discrete.
\end{itemize}
Therefore it is better to preserve the classification up to isomorphism via the first Chern class in $H^{p+2}(X, \mathbb{Z})$: we then introduce a suitable variant of $\check{\rm{C}}$ech cohomology, in order to restore the symmetry even for the cohomology of $\underline{U}(1)$, leading to a good definition of $p$-gerbes with $U(1)$-band also for $p < 0$.

\section{Hat-cohomology}\label{HatCohomology}

We introduce a variant of $\check{\rm{C}}$ech cohomology, starting from sheaves of functions with value in a fixed abelian Lie group $G$, even of dimension $0$ as $\mathbb{Z}$, and then generalizing it to any sheaf of abelian Lie groups, or even topological groups.\footnote{We remark that we do not construct a cohomology theory of sheaves in the ordinary sense, since the canonical axioms are not satisfied \cite{Bredon}.} We denote this cohomology theory as $\hat{H}^{\bullet}(X, \mathcal{F})$ for $\mathcal{F}$ a sheaf of groups, and we denote by $\underline{G}$ the sheaf of smooth $G$-valued functions.

\subsection{Sheaves of $G$-valued functions}

\subsubsection{First attempt}

We keep in mind the case of $(-1)$-gerbes. Since we want to define them as homotopy classes of functions, we require that $\hat{H}^{0}(X, \underline{U}(1))$ is isomorphic to $\check{H}^{0}(X, \underline{U}(1)) \,/\, \textnormal{homotopy}$. Hence for a generic sheaf $\underline{G}$, remembering that $\check{H}^{0}(X, \underline{G})$ is the set of smooth functions $f: X \rightarrow G$, we look for cohomology groups $\hat{H}^{\bullet}(X, \underline{G})$ satisfying:
\begin{equation}\label{HatHFirstAtt}
\begin{array}{l}
	\hat{H}^{0}(X, \underline{G}) \simeq \check{H}^{0}(X, \underline{G}) \,/\, \textnormal{homotopy}\\
	\hat{H}^{i}(X, \underline{G}) \simeq \check{H}^{i}(X, \underline{G}) \; \forall i \geq 1.
\end{array}
\end{equation}
This means that we want to consider as $0$-coboundaries the functions homotopic to the constant function $1$. Thus, we could modify the $\check{\rm{C}}$ech complex in the following way:
\begin{equation}\label{HatCFirstAtt1}
\begin{array}{l}
	\hat{C}^{-1}(X, \underline{G}) = \{f: X \rightarrow G \,|\, f \simeq 1\}\\
	\hat{C}^{i}(X, \underline{G}) = \check{C}^{i}(X, \underline{G}) \; \forall i \geq 0\\ \\
	\hat{\delta}^{-1}(f) = f\\
	\hat{\delta}^{i} = \check{\delta}^{i} \; \forall i \geq 0.
\end{array}
\end{equation}
In this way the corresponding cohomology groups $\hat{H}^{\bullet}(X, \underline{G})$ satisfy exactly \eqref{HatHFirstAtt}, since functions homotopic to $1$ becomes coboundaries by definition. Is this definition correct? Actually, this is not the case. Let us consider the exact sequence:
\begin{equation}\label{ExSeqZRS12}
	0 \longrightarrow \mathbb{Z} \longrightarrow \underline{\mathbb{R}} \longrightarrow \underline{U}(1) \longrightarrow 1
\end{equation}
and the associated sequence in cohomology starting from degree $-1$:
	\[\hat{H}^{-1}(X, \mathbb{Z}) \longrightarrow \hat{H}^{-1}(X, \underline{\mathbb{R}}) \longrightarrow \hat{H}^{-1}(X, \underline{U}(1)) \longrightarrow \hat{H}^{0}(X, \mathbb{Z}) \longrightarrow \hat{H}^{0}(X, \underline{\mathbb{R}}).
\]
Since $\hat{\delta}^{-1}$ is injective, the only $(-1)$-cocycle is $0$ for any sheaf, therefore the first three groups vanish. Since $\mathbb{Z}$ is discrete, the only $\mathbb{Z}$-valued function homotopic to $0$ is the constant function $0$, so that $\hat{H}^{0}(X, \mathbb{Z}) \simeq \check{H}^{0}(X, \mathbb{Z})$. Moreover, since $\mathbb{R}$ is contractible, every $\mathbb{R}$-valued function is homotopic to zero, so that $\hat{H}^{0}(X, \underline{\mathbb{R}}) \simeq 0$. That's why the previous sequence becomes, supposing for simplicity that $X$ is connected:
	\[0 \longrightarrow 0 \longrightarrow 0 \overset{\beta}\longrightarrow \mathbb{Z} \longrightarrow 0.
\]
This sequence is not exact, otherwise $\beta$ should be an isomorphism. To identify the problem, we recall what happens for the ordinary $\rm\check{C}$ech cohomology. If we consider a short exact sequence of \emph{pre}-sheaves $0 \rightarrow \mathcal{P}_{1} \rightarrow \mathcal{P}_{2} \rightarrow \mathcal{P}_{3} \rightarrow 0$, it immediately follows from the definition that the corresponding sequence of $\rm\check{C}$ech-cochain complexes $0 \rightarrow \check{C}^{\bullet}(X, \mathcal{P}_{1}) \rightarrow \check{C}^{\bullet}(X, \mathcal{P}_{2}) \rightarrow \check{C}^{\bullet}(X, \mathcal{P}_{3}) \rightarrow 0$ is exact, therefore, by a standard homological algebra argument \cite{Hatcher}, we get a long exact sequence in cohomology. Then, for a paracompact space (a smooth manifold is paracompact), one shows that even an exact sequence of sheaves $0 \rightarrow \mathcal{F}_{1} \rightarrow \mathcal{F}_{2} \rightarrow \mathcal{F}_{3} \rightarrow 0$ produces a long exact sequence in cohomology \cite{Bredon}. In fact, on such a space, if $\mathcal{P}$ is a pre-sheaf and $\mathcal{P}^{\natural}$ its sheafification, the $\rm\check{C}$ech cohomology groups of $\mathcal{P}$ and $\mathcal{P}^{\natural}$ are canonically isomorphic. Hence, if we consider the quotient \emph{pre}-sheaf $\mathcal{F}_{2}/\mathcal{F}_{1}$, the cohomology of $\mathcal{F}_{3}$ is isomorphic to the one of $\mathcal{F}_{2}/\mathcal{F}_{1}$, so that the long sequence resulting from $0 \rightarrow \mathcal{F}_{1} \rightarrow \mathcal{F}_{2} \rightarrow \mathcal{F}_{3} \rightarrow 0$ is the same that one would get from $0 \rightarrow \mathcal{F}_{1} \rightarrow \mathcal{F}_{2} \rightarrow \mathcal{F}_{2}/\mathcal{F}_{1} \rightarrow 0$, the latter being an exact sequence of pre-sheaves. Therefore, the long sequence is exact. In our case, we have not defined the $(-1)$-degree cochains for a pre-sheaf (actually, we will define them in such a way that they are equal to the ones of the sheafification), hence we work directly with sheaves. In particular, we must verity that, given an exact sequence of abelian groups $0 \rightarrow A \rightarrow B \rightarrow C \rightarrow 0$, the associated sequence of cochains $0 \rightarrow \hat{C}^{-1}(X, \underline{A}) \rightarrow \hat{C}^{-1}(X, \underline{B}) \rightarrow \hat{C}^{-1}(X, \underline{C}) \rightarrow 0$ is exact. With the present definition this is not the case. In fact, let us consider the sequence in degree $-1$ associated to \eqref{ExSeqZRS12}, i.e.:
\begin{equation}\label{SeqChains}
	0 \longrightarrow \hat{C}^{-1}(X, \mathbb{Z}) \longrightarrow \hat{C}^{-1}(X, \mathbb{R}) \longrightarrow \hat{C}^{-1}(X, \underline{U}(1)) \longrightarrow 1
\end{equation}
which is equal to:
\[0 \longrightarrow 0 \longrightarrow \{f: X \rightarrow \mathbb{R}\} \overset{\exp(2\pi i \,\cdot)}\longrightarrow \{f: X \rightarrow \underline{U}(1) \,\vert\, f \simeq 1\} \longrightarrow 0 \; .
\]
It is not exact, since $\exp$ is not injective. A constant function $f: X \rightarrow \mathbb{R}$ with integer value belongs to the kernel of $\exp$, but it is not necessarily $0$. Such a constant function is of course liftable to a $\mathbb{Z}$-valued function, but, in order to lie in $\hat{C}^{-1}(X, \mathbb{Z})$, the lift must be homotopic to the identity, which is the case only for the constant function $0$. The problem is that a homotopy $F: X \times I \rightarrow \mathbb{R}$ between $f$ and $0$ is not liftable to $\mathbb{Z}$, since it assumes continuous values, so, although a non-vanishing integral function $f$ is homotopic to zero as a real function, it is not homotopic to zero as an integral function, coherently with the fact that $\hat{C}^{-1}(X, \mathbb{Z}) = 0$.

\paragraph{}From the previous discussion we deduce that the lack of injectivity of $\exp$ is due to the fact that it destroys the information about the homotopy. Therefore, one possible solution is to include the homotopy itself in the definition. We can define, for $I = [0,1]$:
\begin{equation}\label{HatCFirstAtt2}
\begin{array}{l}
	\hat{C}^{-1}(X, \underline{G}) = \{f: X \times I \rightarrow G \textnormal{ such that } f\vert_{{X \times \{0\}}} = 1\} \\
	\hat{C}^{i}(X, \underline{G}) = \check{C}^{i}(X, \underline{G}) \; \forall i \geq 0\\ \\
	\hat{\delta}^{-1}(f) = f\vert_{{X \times \{1\}}} \\
	\hat{\delta}^{i} = \check{\delta}^{i} \; \forall i \geq 0 \; .
\end{array}
\end{equation}
In this way the sequence \eqref{SeqChains} becomes:
\[0 \longrightarrow 0 \longrightarrow \{f: X \times I \rightarrow \mathbb{R}, f\vert_{{X \times \{0\}}} = 0\} \overset{\exp(2\pi i \,\cdot)}\longrightarrow \{f: X \times I \rightarrow U(1), f\vert_{{X \times \{0\}}} = 1\} \longrightarrow 1.
\]
Now $\exp$ is bijective, as immediately follows from the homotopy lifting property \cite{Hatcher}, since $\mathbb{R}$ is a cover of $U(1)$. We can see directly the injectivity: in fact, let us suppose that $e^{2 \pi i f} = 1$. Then $f$ is integral, thus constant, and, since $f\vert_{{X \times \{0\}}} = 0$, it follows that $f = 0$.

This definition solves the problems about $(-1)$-gerbes, since $\hat{H}^{0}(X, \underline{\mathbb{R}}) = 0$. In fact, if we consider a cocycle $f: X \rightarrow \mathbb{R}$, it is homotopic to the $0$-map being $\mathbb{R}$ contractible, for example via the homotopy $F(x, t) = t f(x)$. Thus, $F \in \hat{C}^{-1}(X, \underline{\mathbb{R}})$ and $\hat{\delta}^{-1}F = f$, so that every $f$ is a coboundary. This means that the Bockstein map $\beta_{0}: \hat{H}^{0}(X, \underline{U}(1)) \rightarrow \hat{H}^{1}(X, \mathbb{Z})$ is now an isomorphism.

\paragraph{}Let us now see what happens for $(-2)$-gerbes. In this case, we have the same problem that we had about $(-1)$-gerbes with ordinary $\check{\rm{C}}$ech cohomology: by definition $\hat{H}^{-1}(X, \underline{\mathbb{R}})$ is the set of real-valued homotopies between $0$ and itself, and this group is not zero, thus the Bockstein map $\beta_{-1}: \hat{H}^{-1}(X, \underline{U}(1)) \rightarrow \hat{H}^{0}(X, \mathbb{Z})$ is not injective. In fact, for $X$ connected, $\hat{H}^{0}(X, \mathbb{Z}) = \mathbb{Z}$ as for $\check{\rm{C}}$ech cohomology; instead, $\hat{H}^{-1}(X, \underline{U}(1))$ is the set of homotopies $f: X \times I \rightarrow U(1)$ between $1$ and itself, and this group is much larger than $\mathbb{Z}$ (it is continuous), unless we consider also $f$ up to homotopy.  What we need is that $\hat{H}^{i}(X, \underline{\mathbb{R}}) = 0$ for every $i$. The right solution is to quotient out also the $(-1)$-cocycles up to homotopy, so that, being $\mathbb{R}$ contractible, even the cohomology in degree $-1$ vanishes. Thus, we can define $\hat{C}^{-2}(X, \underline{G})$ considering homotopies of homotopies, and so on for any degree.

\subsubsection{The correct definition}

We use the notation $I := [0,1]$ and $J^{i} \subset I^{i}$ defined as $J^{i} := I^{i-1} \times \{0\} \cup \partial(I^{i-1}) \times I$ (in particular $J^{1} = \{0\}$, thinking of $I^{0}$ as a point). We define:
\begin{equation}\label{HatCFirst}
\begin{split}
	&\hat{C}^{i}(X, \underline{G}) = \check{C}^{i}(X, \underline{G}) \quad \forall i \geq 0 \\
	&\hat{C}^{-i}(X, \underline{G}) = \{ f: X \times I^{i} \rightarrow G \textnormal{ such that } f\vert_{X \times J^{i}} = 1\} \quad \forall i \geq 1
\end{split}
\end{equation}
with coboundaries:
\begin{equation}\label{HatCCoboundaries}
\begin{split}
	&\hat{\delta}^{i} = \check{\delta}^{i} \quad \forall i \geq 0 \\
	&\hat{\delta}^{-i}(f) = f\vert_{X \times I^{i-1} \times \{1\}} \quad \forall i \geq 1.
\end{split}
\end{equation}
First of all, let us now verify that, given an exact sequence of abelian groups:
\begin{equation}\label{ExSeqABC}
	1 \longrightarrow A \overset{i}\longrightarrow B \overset{\pi}\longrightarrow C \longrightarrow 1
\end{equation}
we obtain an exact sequence of cochain complexes at negative degrees:
\begin{equation}\label{ExSeqABCCochains}
	1 \longrightarrow \hat{C}^{<0}(X, \underline{A}) \overset{i^{*}}\longrightarrow \hat{C}^{<0}(X, \underline{B}) \overset{\pi^{*}}\longrightarrow \hat{C}^{<0}(X, \underline{C}) \longrightarrow 1.
\end{equation}
We verify the exactness in each of the three positions of the sequence:
\begin{itemize}
	\item \emph{Injectivity of $i^{*}$.} Let us fix $f \in \hat{C}^{-i}(X, \underline{A})$. Then, $i^{(-i)}f = i \circ f$ for $i$ defined in \eqref{ExSeqABC}, hence, being $i$ injective, also $i^{(-i)}$ is injective.
	\item \emph{Exactness in the middle.} Since $\pi^{(-i)}f = \pi \circ f$ for $\pi$ defined in \eqref{ExSeqABC}, from the exactness of \eqref{ExSeqABC} in the middle it follows that $\pi^{(-i)}f = 1$ if and only if $f$ can be lifted to $\hat{C}^{-i}(X, \underline{A})$.
	\item \emph{Surjectivity of $\pi^{*}$.} This easily follows from the fact that, for Lie groups, the projection on the homogeneous space $\pi: B \rightarrow B/A$ is an $A$-principal bundle, therefore it has the homotopy lifting property \cite{Hatcher}. Thus, a homotopy $F: X \times I^{i-1} \times I \rightarrow C \simeq B/A$ can be lifted to a homotopy $F: X \times I^{i-1} \times I \rightarrow B$, once that we fix the lift of the starting function, which is $1: X \times I^{i-1} \times \{0\} \rightarrow B$ in the present case. Moreover, it is always possible to choose the lifted homotopy in such a way that it is still relative to $X \times \partial I^{i-1}$: in fact, the homotopy lifting property is by definition the lift extension property \cite{Hatcher} for the couple $(X \times I^{i}, X \times I^{i-1} \times \{0\})$, while its version relative to $X \times \partial I^{i-1}$ is the lift extension property for the couple $(X \times I^{i}, X \times I^{i-1} \times \{0\} \cup X \times \partial I^{i-1} \times I)$; since the couples $(I^{i}, I^{i-1} \times \{0\})$ and $(I^{i}, I^{i-1} \times \{0\} \cup \partial I^{i-1} \times I)$ are homeomorphic, the two properties are equivalent.
\end{itemize}
We can now show the structure of cocycles and coboundaries:
\begin{itemize}
	\item a $(-i)$-cocycle is a function $f: X \times I^{i} \rightarrow G$ such that $f\vert_{X \times \partial I^{i}} = 1$; a $0$-cocycle is a function $f: X \rightarrow G$ as for the $\check{\rm{C}}$ech cohomology;
	\item a $(-i)$-coboundary is a cocycle $f$ homotopic to the identity relatively to $X \times \partial I^{i}$, since, if $F$ is such a homotopy between $f$ and $1$, then $f = \hat{\delta}^{-i-1}F$; for the same reason, a $0$-coboundary is a function $f: X \rightarrow G$ homotopic to $1$.
\end{itemize}
It follows that the hat-cohomology groups are:
\begin{equation}\label{HatCohomologyGroups}
\begin{array}{l}
	\hat{H}^{i}(X, \underline{G}) = \check{H}^{i}(X, \underline{G}) \; \forall i \geq 1\\
	\hat{H}^{0}(X, \underline{G}) = \check{H}^{0}(X, \underline{G}) \,/\, \textnormal{homotopy} \\
	\hat{H}^{-i}(X, \underline{G}) = \{f: X \times I^{i} \rightarrow G,\, f\vert_{X \times \partial I^{i}} = 1\} \,/\, \textnormal{homotopy relative to $\partial I^{i}$} \; \forall i \geq 1.
\end{array}
\end{equation}
Since $I^{i} / \partial I^{i}$ is homeomorphic to the sphere $S^{i}$, with a marked point $p_{0}$ corresponding to $\partial I^{i} / \partial I^{i}$, we can think of $\hat{H}^{-i}(X, \underline{G})$ as the set of functions from $X \times S^{i}$ to $G$ which are equal to $1$ on $X \times \{p_{0}\}$, up to homotopy relative to $X \times \{p_{0}\}$. In particular, if we call $F(X, G)$ the space of smooth functions from $X$ to $G$, it follows that:
	\[\hat{H}^{-i}(X, \underline{G}) \simeq \pi_{i}(F(X, G), 1) \; \forall i \geq 1,
\]
where $\pi_{i}(F(X, G), 1)$ is the $i$-th homotopy group with base-point the constant function $1$. Hence, the long exact sequence in cohomology at negative degrees, associated to an exact sequence of groups $0 \rightarrow A \rightarrow B \rightarrow C \rightarrow 0$, is the long exact sequence in homotopy associated to the fibration $F(X, A) \rightarrow F(X, B) \rightarrow F(X, C)$ \cite{Hatcher}.

\subsection{Generalization}\label{Generalization}

We can generalize the definition of hat-cohomology to any sheaf of topological groups. In particular, we remark that a homotopy between two functions $f,g: X \rightarrow Y$ is a continuous (or smooth) path in the space of continuous (or smooth) functions from $X$ to $Y$: in other words, a homotopy $F: X \times I \rightarrow Y$, can be thought of as a path $\varphi: I \rightarrow C^{0}(X,Y)$ such that $\varphi(0) = f$ and $\varphi(1) = g$, defined by $\varphi(t)(x) = F(t, x)$. For the case $Y = G$, where $G$ is a Lie group or topological group, it holds that $C^{0}(X, G) = \underline{G}(X)$, where $\underline{G}(X)$ is the group of global sections of the sheaf $\underline{G}$. Therefore, we can generalize \eqref{HatCFirst} to a generic sheaf of topological groups $\mathcal{F}$ in the following way:
\begin{equation}\label{HatCGeneral}
\begin{split}
	&\hat{C}^{i}(X, \mathcal{F}) = \check{C}^{i}(X, \mathcal{F}) \quad \forall i \geq 0\\
	&\hat{C}^{-i}(X, \mathcal{F}) = \{ \varphi: I^{i} \rightarrow \mathcal{F}(X) \textnormal{ such that } \varphi\vert_{J^{i}} = e\} \quad \forall i \geq 1
\end{split}
\end{equation}
where $e$ is the unit of $\mathcal{F}(X)$, with coboundaries:
\begin{equation}\label{HatCGeneralCoboundaries}
\begin{split}
	&\hat{\delta}^{i} = \check{\delta}^{i} \quad \forall i \geq 0 \\
	&\hat{\delta}^{-i}(\varphi) = \varphi\vert_{I^{i-1} \times \{1\}} \quad \forall i \geq 1.
\end{split}
\end{equation}
It follows that:\footnote{In the following equalities involving homotopy groups, we use the fact that, for a topological group $A$, the multiplication on $\pi_{n}(A,e)$, as a homotopy group, coincides with the pointwise multiplication of functions from $(S^{n}, *)$ to $(A, e)$, up to homotopy relative to $*$ \cite{Hu}.}
\begin{equation}\label{HatCGeneralCohomology}
\begin{split}
	&\hat{H}^{i}(X, \mathcal{F}) = \check{H}^{i}(X, \mathcal{F}) \quad \forall i \geq 1\\
	&\hat{H}^{0}(X, \mathcal{F}) = \pi_{0}(\mathcal{F}(X)) \\
	&\hat{H}^{-i}(X, \mathcal{F}) = \pi_{i}(\mathcal{F}(X),e) \quad \forall i \geq 1.
\end{split}
\end{equation}
For pre-sheaves, we can use the same definition replacing $\mathcal{F}(X)$ with $\check{H}^{0}(X, \mathcal{F})$, which coincides up to isomorphism with $\mathcal{F}^{\natural}(X)$, for $\mathcal{F}^{\natural}$ the sheafification of $\mathcal{F}$ \cite{Bredon}. The problem with this generalization is to find the correct hypotheses in order to verify the exactness of \eqref{ExSeqABCCochains}, but we do not develop this problem here, since it is not needed in the following.

\subsection{$p$-Gerbes and hat-cohomology}

Let us now come back to the exact sequence:
	\[0 \longrightarrow \mathbb{Z} \longrightarrow \underline{\mathbb{R}} \longrightarrow \underline{U}(1) \longrightarrow 1.
\]
First of all we obtain the desired result that $\hat{H}^{i}(X, \underline{\mathbb{R}}) = 0 \, \forall i \in \mathbb{Z}$, so that $\beta^{i}: \hat{H}^{i}(X, \underline{U}(1)) \rightarrow \hat{H}^{i+1}(X, \mathbb{Z})$ is an isomorphism for every $i \in \mathbb{Z}$. Thus, we can give a good topological definition of abelian $p$-gerbe for every $p \in \mathbb{Z}$, in such a way that there is a natural bijection between isomorphism classes of $p$-gerbes and elements of $\hat{H}^{i+1}(X, \underline{U}(1))$. In particular, considering the remark after equation \eqref{HatCohomologyGroups}, we define:
\begin{Def} For $p < 0$:
\begin{itemize}
	\item a \emph{$p$-gerbe with $U(1)$-band} is a function $f: X \times S^{-p-1} \rightarrow U(1)$ such that $f\vert_{X \times \{p_{0}\}} = 1$;
	\item an \emph{isomorphism} between two $p$-gerbes $f,g: X \times S^{-p-1} \rightarrow U(1)$ is a homotopy $F: X \times S^{-p-1} \times I \rightarrow U(1)$ between $f$ and $g$ relative to $X \times \{p_{0}\}$;
	\item a $p$-gerbe is \emph{trivial} if it is isomorphic to the constant function $1$.
\end{itemize}
\end{Def}
For $p = -1$ the definition leads to a function $f: X \times S^{0} \rightarrow U(1)$ such that $f\vert_{X \times \{p_{0}\}} = 1$, but, since $S^{0}$ is the disjoint union of two points, one of which is $p_{0}$, such a function is equivalent to a function $f: X \rightarrow U(1)$ without restrictions, therefore we recover the previous definition. The first Chern class of a $p$-gerbe with $U(1)$-band belongs to $H^{i+2}(X, \mathbb{Z})$, and it completely determines the $p$-gerbe up to isomorphism. Actually, the previous definition can be easily generalized to the non-abelian case, replacing $U(1)$ with $U(n)$, and, via the suspension isomorphism, we can define the Chern classes and the Chern characters of a non-abelian $p$-gerbe for $p < 0$. Moreover, there are non-trivial non-abelian $p$-gerbes for any $p \in \mathbb{Z}$. We do not develop this topic further, since abelian gerbes are sufficient in order to describe Ramond-Ramond fields. Let us now classify the isomorphism classes of abelian $p$-gerbes for $p \leq -1$:
\begin{itemize}
	\item For $p = -1$, as we have already said, isomorphism classes of $(-1)$-gerbes are homotopy classes of functions with value in $U(1)$, classified by $H^{1}(X, \mathbb{Z})$.
	\item For $p = -2$, isomorphism classes of $(-2)$-gerbes are elements of $\hat{H}^{-1}(X, \underline{U}(1))$, i.e.\ homotopies between $1: X \rightarrow U(1)$ and itself, up to homotopy relative to $X \times \{0,1\}$. They are classified by $H^{0}(X, \mathbb{Z})$, which is isomorphic to $\mathbb{Z}$ for $X$ connected. What is the explicit correspondence between classes of homotopies from $1$ to itself and $\mathbb{Z}$, i.e.\ the Bockstein map $\beta^{(-1)}$? Given a representative $f$, we lift it from $U(1)$ to $\mathbb{R}$: we obtain a homotopy between $1$ and another lifting of $1$ in $\mathbb{R}$, which is an integer number $n \in \mathbb{Z}$. It is easy to see from the theory of coverings \cite{Hatcher} that $n$ depends only on the homotopy class of $f$ relative to the boundary, so that we obtain a well defined map $\beta^{(-1)}f = n$.
	\item Since abelian $p$-gerbes are classified by $H^{p+2}(X, \mathbb{Z})$, \emph{any $p$-gerbe for $p \leq -3$ is trivial}.
\end{itemize}
We thus obtain a picture suitable for type II superstring theory. In fact, as we have discussed in section \ref{RRpGerbes}, we need non-trivial $p$-gerbes exactly for $p \geq -2$, and it is what naturally comes from the definition of hat-cohomology.

\section{Differential geometry and hat-cohomology}\label{DiffGeometry}

Up to now we have discussed only the topology of $p$-gerbes via hat-cohomology. We now study their differential geometry, discussing connections with associated curvature and holonomy.

\subsection{Connections, holonomy and curvature}

For $p \geq 0$ a $p$-gerbe with connection corresponds up to isomorphism to an element of the Deligne cohomology group $\check{H}^{p+1}(X, \underline{U}(1) \rightarrow \Omega^{1}_{\mathbb{R}} \rightarrow \cdots \rightarrow \Omega^{p+1}_{\mathbb{R}})$. Let us start with $(-1)$-gerbes: according to this definition, a $(-1)$-gerbe with connection corresponds to an element of $\check{H}^{0}(X, \underline{U}(1))$. This definition seems correct. In fact, as we have seen, a $(-1)$-gerbe is a function with value in $U(1)$, and its class up to isomorphism is the homotopy class of the function. Considering that, for $p \geq 0$, the local potentials defining a connection are real-valued forms, and that the local potentials for $p = -1$ must be $0$-forms, we can give the following definition:
\begin{Def} The only connection on a $(-1)$-gerbe $f: X \rightarrow U(1)$ is the function $C_{0}: X \rightarrow \mathbb{R}/\mathbb{Z}$ such that $f = e^{2\pi i C}$, which can be represented by local potentials $(C_{0})_{\alpha}: U_{\alpha} \rightarrow \mathbb{R}$ with respect to a good cover $\mathfrak{U} = \{U_{\alpha}\}_{\alpha \in I}$.
\end{Def}
The holonomy of such a connection must be a well-defined function from the space of $0$-submanifolds of $X$ to $U(1)$. Since the connected $0$-submanifolds are the points (and it is not restrictive to consider connected submanifolds), the holonomy must be a function from $X$ to $U(1)$, i.e.\ the $(-1)$-gerbe $f$ itself. Thus, \emph{the holonomy of a connection on a $(-1)$-gerbe is the gerbe itself, or equivalently the choice of a representative within the isomorphism class}. Two homotopic functions are isomorphic as topological $(-1)$-gerbes, but not geometrically as gerbes with connection.

Let us now discuss the curvature. For $p \geq 0$ it is defined considering the highest degree forms of the connection and computing the exterior derivative. For $(-1)$-gerbes, this is equivalent to consider the differential $dC_{0}$ for $C_{0}: X \rightarrow \mathbb{R}/\mathbb{Z}$ the connection. Since $f = e^{2\pi i C_{0}}$, it follows that $dC_{0} = \frac{1}{2\pi i} d\log f = \frac{1}{2\pi i} f^{-1}df$. This is consistent, since, if we consider the complex $\underline{U}(1) \rightarrow \Omega^{1}_{\mathbb{R}} \rightarrow \cdots \rightarrow \Omega^{p+1}_{\mathbb{R}}$, the first boundary is $\tilde{d} = \frac{1}{2\pi i}d \circ \log$ instead of $d$, thus, by analogy with the case $p \geq 0$, it is also natural to define the curvature of a $(-1)$-gerbe as $\tilde{d}f = \frac{1}{2\pi i} f^{-1}df$. One can prove that the de-Rahm cohomology class $[\frac{1}{2\pi i} f^{-1}df]_{dR}$ corresponds to $c_{1}([f]) \otimes_{\mathbb{Z}} \mathbb{R}$ under the standard isomorphism $H^{1}_{dR}(X) \simeq H^{1}(X, \mathbb{R})$.

For $(-2)$-gerbes there is no holonomy, since it should be defined on the space of $(-1)$-submanifolds. For what concerns the curvature, it must be a closed integral $0$-form, i.e.\ a constant integral function $F$, such that $[F]_{dR} \simeq c_{1}(\mathcal{G}) \otimes_{\mathbb{Z}} \mathbb{R}$. Since $c_{1}(\mathcal{G}) \in \hat{H}^{0}(X, \mathbb{Z}) \simeq \mathbb{Z}$, we deduce that the curvature of a $(-2)$-gerbe is simply its first Chern class.

\paragraph{Remark:} It seems natural to extend the definition of hat-cohomology to the one of hat-hypercohomology, in order to describe connections and holonomies in this way. Actually, this extension can be defined, but it gives no interesting information, therefore, in order to deal with connections, it is better to use Deligne cohomology as we did up to now. The details can be found in the appendix \ref{HatHyperCohomology}.

\paragraph{}For example, if we start with the trivial background $X = \mathbb{R}^{1,9}$, every function $f: X \rightarrow U(1)$ is homotopic to the constant function $1$, since the domain is contractible. Thus $\hat{H}^{0}(X, \underline{U}(1)) = 0$ as we expect, so that every $(-1)$-gerbe is topologically trivial; this also follows from the fact that $H^{1}(X, \mathbb{Z}) = 0$ and $\hat{H}^{0}(X, \underline{U}(1)) \simeq H^{1}(X, \mathbb{Z})$. A connection on the trivial gerbe is determined by its holonomy $\tilde{C}_{0}: \mathbb{R}^{1,9} \rightarrow U(1)$. We can choose only one patch, coinciding with the whole $X$, and a unique Ramond-Ramond potential $C_{0}: X \rightarrow \mathbb{R}$ up to global gauge transformations $C_{0} \rightarrow C_{0} + n$ with $n \in \mathbb{Z}$: then $G_{1} = dC_{0}$ is the field strength and $\tilde{C}_{0} = e^{2\pi i C_{0}}$ is the holonomy. Thus, all the $(-1)$-gerbes are topologically trivial, and they are geometrically classified by a function $\tilde{C}_{0}: \mathbb{R}^{1,9} \rightarrow U(1)$, or, equivalently, by an equivalence class of functions $C_{0}: X \rightarrow \mathbb{R}$ up to the addition of an integer.

\paragraph{Remark:} We see even in this trivial example that in the gauge transformations $C_{0} \rightarrow C_{0} + n$ of the $0$-degree Ramond-Ramond potentials it is necessary to impose the constraint that $n \in \mathbb{Z}$, otherwise the holonomy $\tilde{C}_{0} = e^{2\pi i C_{0}}$, which is the path-integral measure corresponding to the Wess-Zumino action for a D$(-1)$-brane, should not be well-defined. This seems to be in contrast with the gauge transformations of the higher-order potentials, which are $C_{p} \rightarrow C_{p} + d\Lambda_{p-1}$, without integrality conditions. We will discuss later the reason of this apparent inconsistency. $\square$

\paragraph{}If we compactify one dimension, so that $X = \mathbb{R}^{1,8} \times U(1)$, since $X$ retracts on $U(1)$, the homotopy classes of functions $f: X \rightarrow U(1)$ coincide with the homotopy classes of functions $\varphi: U(1) \rightarrow U(1)$, in particular the class of $f$ corresponds to the class of $\varphi$ defined by $\varphi(u) := f(0, u)$. Thus $\hat{H}^{0}(X, \underline{U}(1)) \simeq \pi_{1}(U(1)) \simeq \mathbb{Z}$: this also follows from the fact that $H^{1}(X, \mathbb{Z}) \simeq \mathbb{Z}$ and $\hat{H}^{0}(X, \underline{U}(1)) \simeq H^{1}(X, \mathbb{Z})$. It is easy to find one representative for each equivalence class in $\hat{H}^{0}(X, \underline{U}(1))$: for $n \in \mathbb{Z}$, we consider the function $\tilde{C}_{0}(x, z) = z^{n}$, whose class corresponds to $n$ under the isomorphism $\hat{H}^{0}(X, \underline{U}(1)) \simeq \pi_{1}(U(1)) \simeq \mathbb{Z}$, as the reader can verify. This classifies the $(-1)$-gerbes topologically. For a given superstring background, with local Ramond-Ramond potentials $(C_{0})_{\alpha}$ with respect to a fixed cover, one considers the globally defined function $\tilde{C}_{0} = e^{2\pi i (C_{0})_{\alpha}}$, which is both the connection and the holonomy, that classifies the $(-1)$-gerbe geometrically. Then, to catch the topological information, which is the equivalence class up to homotopy, one computes the first Chern class as shown above: if the latter corresponds to $n$ under the isomorphism $H^{1}(X, \mathbb{Z}) \simeq \mathbb{Z}$, then the given $\tilde{C}_{0}$ will be homotopic to $f(x, z) = z^{n}$.

If we compactify two dimensions, so that $X = \mathbb{R}^{1,7} \times T^{2}$, we get that $\hat{H}^{0}(X, \underline{U}(1)) \simeq H^{1}(X, \mathbb{Z}) \simeq \mathbb{Z} \oplus \mathbb{Z}$, thus the $(-1)$-gerbes are topologically classified by this group. One representative of the class corresponding to $(n, m) \in \mathbb{Z} \oplus \mathbb{Z}$ is $\tilde{C}_{0}(x,z, w) = z^{n}w^{m}$ for $x \in \mathbb{R}^{1,7}$ and $z,w \in U(1)$. For other backgrounds the way of arguing is the same, considering the topology of the internal manifold when computing the cohomology groups or finding a suitable good cover for the local potentials.

\subsection{Flat connections}

For a $p$-gerbe with $p \geq 0$, a connection is flat when its curvature is $0$. In this case, the holonomy is quantized and, for $p \geq 0$, flat $p$-gerbes are classified by $H^{p+1}(X, U(1))$, for $U(1)$ the \emph{constant} sheaf. What happens for $(-1)$-gerbes? They are classified by $H^{0}(X, U(1))$, i.e.\ by locally constant functions $f: X \rightarrow U(1)$. This is correct since, for a function $f: X \rightarrow U(1)$ thought of as $(-1)$-gerbe with connection, the curvature $\frac{1}{2\pi i} f^{-1}df$ is zero if and only if $df = 0$, i.e.\ $f$ is locally constant. In this case the holonomy, which is $f$ itself, is quantized and depends only on the connected component of the domain. We can make a remark: any locally constant function is homotopic to $1$, since $U(1)$ is pathwise connected (to realize a homotopy it is enough to consider for each component a path in $U(1)$ from the value of $f$ to $1$). Thus, any flat $(-1)$-gerbe is topologically trivial. In particular, $H^{1}(X, \mathbb{Z})$ cannot have torsion. This is a topological fact: the more general way to prove it is to use universal coefficient theorem \cite{Hatcher}, which implies that $H^{1}(X, \mathbb{Z}) \simeq \Hom(H_{1}(X, \mathbb{Z}), \mathbb{Z})$, so that it cannot have torsion. But we can prove it also in a more concrete way. In fact, since $H^{i}(X, \mathbb{R}) \simeq H^{i}(X, \mathbb{Z}) \otimes_{\mathbb{Z}} \mathbb{R}$, it follows that $H^{i}(X, \mathbb{Z})$ has no torsion if and only if the natural map $\iota: H^{i}(X, \mathbb{Z}) \rightarrow H^{i}(X, \mathbb{R})$ is injective. We have a long exact sequence in $\check{\rm{C}}$ech cohomology of constant sheaves:
	\[\check{H}^{0}(X, \mathbb{Z}) \longrightarrow \check{H}^{0}(X, \mathbb{R}) \longrightarrow \check{H}^{0}(X, U(1)) \longrightarrow \check{H}^{1}(X, \mathbb{Z}) \longrightarrow \check{H}^{1}(X, \mathbb{R})
\]
which is isomorphic to:
	\[\mathbb{Z} \longrightarrow \mathbb{R} \overset{\exp}\longrightarrow U(1) \overset{\beta}\longrightarrow \check{H}^{1}(X, \mathbb{Z}) \overset{\iota}\longrightarrow \check{H}^{1}(X, \mathbb{R}).
\]
The map $\exp$ is surjective, thus, by exactness, $\beta = 0$ and, again by exactness, $\iota$ is injective. Any flat $(-1)$-gerbe is topologically trivial, and, in fact, it can be represented by a global constant function $f$, i.e.\ it can always be lifted to a globally defined potential belonging to $\check{H}^{0}(X, \mathbb{R})$.

For $(-2)$-gerbes the situation is very simple. Since they are classified by $\mathbb{Z}$, the only flat gerbe is the trivial one. Since the curvature completely determines the gerbe, this is not a surprise.

\subsection{Gauge transformations of $C_{0}$}\label{GaugeC0}

For the gauge transformations of $C_{0}$, which we write as $(C_{0})_{\alpha} - (C_{0})_{\beta} = n_{\alpha\beta}$, one must impose $n_{\alpha\beta} \in \mathbb{Z}$, otherwise the holonomy $\tilde{C}_{0} = e^{2\pi i C_{0}}$ is not a well-defined function. Let us consider the gauge transformations of the highest degree potentials for $p \geq 0$: they are of the form $(C_{p+1})_{\alpha} - (C_{p+1})_{\beta} = d(C'_{p})_{\alpha\beta}$. One can ask why the condition on the transition forms is to be integral when the degree is $0$ and to be exact for higher degrees. It seems an unnatural jump. Actually, this can be understood considering the so-called \emph{large gauge transformations}. Let us consider a cover $\mathfrak{U} = \{U_{\alpha}\}_{\alpha \in I}$ which is not a good cover. In particular, let us suppose that the open sets $U_{\alpha}$ are contractible, but that there exists a double intersection $U_{\alpha\beta}$ such that $H_{p+1}(U_{\alpha\beta}, \mathbb{Z}) \neq 0$. In this case:
	\[\begin{split}
	&(G_{p+2})\vert_{U_{\alpha}} = d(C_{p+1})_{\alpha}\\
	&d \bigl( (C_{p+1})_{\alpha} - (C_{p+1})_{\beta} \bigr) = 0
\end{split}\]
but we cannot deduce that the form $(\Lambda_{p+1})_{\alpha\beta} := (C_{p+1})_{\alpha} - (C_{p+1})_{\beta}$ is exact on $U_{\alpha\beta}$. Let us see that nature of $(\Lambda_{p+1})_{\alpha\beta}$. We can consider a non-trivial $(p+1)$-cycle $\Gamma_{p+1} \subset U_{\alpha\beta}$. Then, since $U_{\alpha}$ and $U_{\beta}$ are contractible, we can find two $(p+1)$-chains $(\Gamma_{p+2})_{\alpha} \subset U_{\alpha}$ and $(\Gamma_{p+2})_{\beta} \subset U_{\beta}$ such that $\partial(\Gamma_{p+2})_{\alpha} = \Gamma_{p+1}$ and $\partial(\Gamma_{p+2})_{\beta} = -\Gamma_{p+1}$. Hence, $\Gamma_{p+2} = (\Gamma_{p+2})_{\alpha} \cup (\Gamma_{p+2})_{\beta}$ is a $(p+2)$-cycle in $U_{\alpha} \cup U_{\beta}$. Since $G_{p+2}$ is integral, $\int_{\Gamma_{p+2}} G_{p+2} \in \mathbb{Z}$. But the following identity holds:
	\[\begin{split}
	\int_{\Gamma_{p+2}} G_{p+2} &= \int_{(\Gamma_{p+2})_{\alpha}} d(C_{p+1})_{\alpha} + \int_{(\Gamma_{p+2})_{\beta}} d(C_{p+1})_{\beta} \\
	& = \int_{\Gamma_{p+1}} \bigl( (C_{p+1})_{\alpha} - (C_{p+1})_{\beta} \bigr) = \int_{\Gamma_{p+1}} (\Lambda_{p+1})_{\alpha\beta}
\end{split}\]
therefore:
	\[\int_{\Gamma_{p+1}} (\Lambda_{p+1})_{\alpha\beta} \in \mathbb{Z}.
\]
This implies that $(\Lambda_{p+1})_{\alpha\beta}$ is an integral form. These kind of gauge transformations, involving non-trivial open subsets, are large gauge transformations. Actually, we can argue in this way only for a cycle $\Gamma_{p+1}$ which, although non-trivial in $U_{\alpha\beta}$, is trivial on the whole space, otherwise it cannot be contained in the contractible chart $U_{\alpha}$ or $U_{\beta}$. In general it is anyway true that $(\Lambda_{p+1})_{\alpha\beta}$ must be integral, otherwise the holonomy along $\Gamma_{p+1}$ would not be well-defined: in fact, being $\Gamma_{p+1}$ contained in the charts $U_{\alpha}$ and $U_{\beta}$, the holonomy can be defined as $\exp(\int_{\Gamma_{p+1}} (C_{p+1})_{\alpha})$ or $\exp(\int_{\Gamma_{p+1}} (C_{p+1})_{\beta})$, and the difference must be zero.

We have thus shown that, for large gauge transformations, the variation is an integral form. If we work with a good cover, as usually happens for smooth manifolds, there are no large gauge transformations. Since the cohomology of a contractible set is trivial at any non-zero degree, and since all of the integral forms must be closed, it follows that the only integral forms are the exact ones. Instead, only in degree $0$ there is a non-trivial cohomology group for contractible sets, in particular constant integral functions are integral non-exact 0-forms, and this explains the apparent difference between the transition functions of $C_{0}$ and the one of higher degree potentials for good covers.

\subsection{Dirac quantization condition in low degree}

Let us analyze the meaning of the Dirac quantization condition for $(-1)$-gerbes and $(-2)$-gerbes. We briefly recall what happens for $p = 0$, the case $p > 0$ being entirely analogous. We consider a closed 2-form $F$ on $X$, and a good cover $\mathfrak{U} = \{U_{\alpha}\}_{\alpha \in I}$ on $X$. We can now apply Poincar\'e lemma iteratively:
\begin{itemize}
	\item since $dF = 0$ and $U_{\alpha}$ is contractible, we can find local 1-forms $\{A_{\alpha}\}_{\alpha \in I}$ such that $F\vert_{U_{\alpha}} = dA_{\alpha}$;
	\item since $dA_{\alpha}\vert_{U_{\alpha\beta}} = dA_{\beta}\vert_{U_{\alpha\beta}}$, we can find local 0-forms $f_{\alpha\beta}$ such that $A_{\alpha}\vert_{U_{\alpha\beta}} - A_{\beta}\vert_{U_{\alpha\beta}} = d f_{\alpha\beta}$;
	\item since $d(f_{\alpha\beta} + f_{\beta\gamma} + f_{\gamma\alpha}) = A_{\alpha} - A_{\beta} + A_{\beta} - A_{\gamma} + A_{\gamma} - A_{\alpha} = 0$, it follows that $f_{\alpha\beta} + f_{\beta\gamma} + f_{\gamma\alpha} = f_{\alpha\beta\gamma}$ constant.
\end{itemize}
In this way realize the class $[F]_{dR} \in H^{2}_{dR}(X)$ as a $\check{\rm{C}}$ech cohomology class $[\{f_{\alpha\beta\gamma}\}] \in \check{H}^{2}(X, \mathbb{R})$. If we consider $g_{\alpha\beta} = e^{2\pi i f_{\alpha\beta}}$, we obtain transition functions with values in $U(1)$ such that $g_{\alpha\beta}g_{\beta\gamma}g_{\gamma\alpha} = g_{\alpha\beta\gamma}$, with $g_{\alpha\beta\gamma}$ constant. If $g_{\alpha\beta\gamma} = 1$ they are transition functions of a line bundle $L$. In this case $f_{\alpha\beta\gamma} \in \mathbb{Z}$ so that they realize a class in $\check{H}^{2}(X, \mathbb{Z})$ which is the first Chern class $c_{1}(L)$. It follows that $F$ is a curvature of the connection $[\{g_{\alpha\beta}, -A_{\alpha}\}] \in \check{H}^{1}(X, \underline{U}(1) \rightarrow \Omega^{1}_{\mathbb{R}})$ and, as we have already said, $c_{1}(L) \otimes_{\mathbb{Z}} \mathbb{R} \simeq [F]_{dR}$. Only in this case holonomy is a well-defined function on the loop space of $X$. Vivecersa, for $F$ integral we can always choose $f_{\alpha\beta} \in \mathbb{Z}$. In fact, since $[\{f_{\alpha\beta\gamma}\}]$ is integral, it follows that $f_{\alpha\beta\gamma} = n_{\alpha\beta\gamma} + \check{\delta}^{2}\{c_{\alpha\beta}\}$ with $c_{\alpha\beta}$ real constant. Hence, $f'_{\alpha\beta} = f_{\alpha\beta} + c_{\alpha\beta} \in \mathbb{Z}$, and, since $df'_{\alpha\beta} = df_{\alpha\beta}$, we can replace $f_{\alpha\beta}$ with $f'_{\alpha\beta}$. This proves that any integral form is the curvature of a connection on a line bundle.

\paragraph{}For $(-1)$-gerbes, let us start from a closed 1-form $F$. Then:
\begin{itemize}
	\item since $dF = 0$ and $U_{\alpha}$ is contractible, we can find local functions $\{f_{\alpha}\}_{\alpha \in I}$ such that $F\vert_{U_{\alpha}} = d f_{\alpha}$;
	\item since $df_{\alpha}\vert_{U_{\alpha\beta}} = df_{\beta}\vert_{U_{\alpha\beta}}$, it follows that $f_{\alpha} - f_{\beta} = f_{\alpha\beta}$ constant.
\end{itemize}
We thus get that $[F]_{dR} \simeq [\{f_{\alpha\beta}\}] \in \check{H}^{1}(X, \mathbb{R})$. If we consider $g_{\alpha} = e^{2\pi i f_{\alpha}}$ we obtain that $g_{\alpha}^{-1}g_{\beta} = g_{\alpha\beta}$ constant. If $g_{\alpha\beta} = 1$ we obtain a global function $g: X \rightarrow U(1)$, which is a $(-1)$-gerbe and coincides with its holonomy. This happens if and only if $f_{\alpha\beta} \in \mathbb{Z}$, and in this case $[F]_{dR}$ is integral and $F = \frac{1}{2\pi i}g^{-1}dg$. Vicevesa one can prove that, for $F$ integral, we can always find $g$ such that $F = \frac{1}{2\pi i}g^{-1}dg$. Thus, as for $p \geq 0$, \emph{the Dirac quantization condition is equivalent for the holonomy to be a globally defined function}. In particular, this means that the Wess-Zumino action of a D-instanton has a well-defined exponential, the latter being the path-integral measure: otherwise, on the intersection of two local charts, such an exponential should depend on the choice of the chart.
	
\paragraph{}Some comments are in order about the quantization of $G_{0}$. Such a field-strength arises from a D$8$-brane, with $9$-dimensional world-volume, which breaks the space-time in two disconnected components. The simplest example is a magnetic monopole in 0+1 space-time dimensions. In this case the space-time is $\mathbb{R}$, and the magnetic monopole with charge $q$ is fixed in the origin. Then, a linking manifold of the charge is the sphere $S^{0}$, i.e.\ the disjoint union of the two points $-1$ and $1$. Therefore, the field-strength is a $0$-form $F$, i.e. a function $F: \mathbb{R} \setminus \{0\} \rightarrow \mathbb{R}$. It must be closed, hence locally constant on $\mathbb{R} \setminus \{0\}$: this means that it assumes a constant value $\alpha^{+}$ on the set of positive numbers and a constant value $\alpha^{-}$ on the set of negative numbers. Then:
	\[q = \int_{S^{0}} F = F(1) - F(-1) = \alpha^{+} - \alpha^{-},
\]
therefore the Dirac quantization conditions means in this case that $\alpha^{+} - \alpha^{-} \in \mathbb{Z}$, up to a normalization constant. It is not necessary that $\alpha^{+}$ and $\alpha^{-}$ are separately integral. In other words, considering the $0$-degree de-Rham cohomology $H^{0}_{dR}(\mathbb{R} \setminus \{0\}) \simeq \mathbb{R} \oplus \mathbb{R}$, it is not necessary that $F$ lies in the integral lattice $H^{0}(\mathbb{R} \setminus \{0\}, \mathbb{Z}) \simeq \mathbb{Z} \oplus \mathbb{Z}$, but only in the subgroup $\{(\alpha^{+}, \alpha^{-}) \,\vert\, \alpha^{+} - \alpha^{-} \in \mathbb{Z}\}$. In string theory, when we consider the field-strength $G_{0}$, we are considering a space-time which should be one of the two halves of a bigger space-time, which is split in two parts by a D$8$-brane. Therefore, it is not strictly necessary that $G_{0}$ is integral: this is true if we assume that, in the other half that we do not consider, it is set to $0$.

\section{Conclusions}\label{Conclusions}

We have shown in this paper how to describe in a unitary way the geometry of Ramond-Ramond field strengths and potentials, including the low-degree cases, and consequently the geometry of D-brane charges and Wess-Zumino action including D-instantons. To achieve this aim, we have introduced a variant of the ordinary $\check{\rm{C}}$ech cohomology, which we have called hat-cohomology. With this language, we can define abelian $p$-gerbes with connection for any $p \in \mathbb{Z}$:
\begin{itemize}
	\item topologically a $p$-gerbe up to isomorphism corresponds to an element of $\hat{H}^{p+1}(X, \underline{U}(1))$;
	\item geometrically a $p$-gerbe with connection up to isomorphism corresponds to an element of $\check{H}^{p+1}(X, \underline{U}(1) \rightarrow \Omega^{1}_{\mathbb{R}} \rightarrow \cdots \rightarrow \Omega^{p+1}_{\mathbb{R}})$.
\end{itemize}
It follows that there exist non-trivial gerbes only for $p \geq -2$. The Ramond-Ramond field strength $G_{p+2}$ is a curvature on a $p$-gerbe for $p \geq -2$, and the Ramond-Ramond potentials $C_{p+1}$ are the local top-forms of a connection on the corresponding $p$-gerbe for $p \geq -1$.

For the case $p = -1$, the field strength is the integral 1-form $G_{1}$, whose holonomy is the globally defined function $\tilde{C}_{0} = e^{2\pi i C_{0}}$, where $C_{0}: X \rightarrow \mathbb{R}/\mathbb{Z}$ is the connection. The topological data, i.e.\ the $(-1)$-gerbe up to isomorphism, is the equivalence class up to homotopy $[\,\tilde{C}_{0}\,] \in \hat{H}^{0}(X, \underline{U}(1))$. The Dirac quantization condition for $G_{1}$ is equivalent to the fact that $\tilde{C}_{0}$ is globally defined, i.e.\ that the Wess-Zumino action for a D-instanton:
	\[S_{WZ} = q_{-1} \cdot C_{0}(WY_{-1})
\]
has a well-defined exponential, the latter being the path-integral measure.

\section*{Acknowledgements}

The author is financially supported by FAPESP (Funda\c{c}\~ao de Amparo \`a Pesquisa do Estado de S\~ao Paulo).

\appendix

\section{Chern class of a function}\label{ChernClassF}

We prove the following lemma, which was stated in subsection \ref{LowDegRR}.

\begin{Lemma} Let us give to $S^{1} \subset \mathbb{C}$ the counter-clockwise orientation, so that we fix an oriented generator $1 \in H^{1}(S^{1}, \mathbb{Z}) \simeq \mathbb{Z}$. Then, for $f: X \rightarrow S^{1}$:
	\[c_{1}(f) = f^{*}(1) \; .
\]
\end{Lemma}
\textbf{Proof:} For $w = e^{\frac{2}{3}i \pi}$, let us consider the three points $1, w, w^{2} \in S^{1}$. We consider on $S^{1}$ the good cover $\mathfrak{U} = \{U_{0}, U_{1}, U_{2}\}$ where $U_{0} = \{e^{2\pi i\theta} \,\vert\, 0 < \theta < \frac{2}{3}\}$, $U_{1} = \{e^{2\pi i\theta} \,\vert\, \frac{1}{3} < \theta < 1\}$ and $U_{2} = \{e^{2\pi i\theta} \,\vert\, \frac{2}{3} < \theta < \frac{4}{3}\}$: in this way, orienting $S^{1}$ counterclockwise, $U_{0} = (1, w^{2})$ and $w \in U_{0}$, $U_{1} = (w, 1)$ and $w^{2} \in U_{1}$, $U_{2} = (w^{2}, w)$ and $1 \in U_{0}$. We now compute the $\rm\check{C}$ech cohomology of $\mathbb{Z}$ with respect to $\mathfrak{U}$. The double intersections are $U_{01}$, $U_{12}$ and $U_{02}$, and there are no higher-order intersections. Thus:
	\[\begin{array}{l}
	\check{C}^{0}(\mathfrak{U}, \mathbb{Z}) = \mathbb{Z} \oplus \mathbb{Z} \oplus \mathbb{Z}\\
	\check{C}^{1}(\mathfrak{U}, \mathbb{Z}) = \mathbb{Z} \oplus \mathbb{Z} \oplus \mathbb{Z}\\
	\check{C}^{i}(\mathfrak{U}, \mathbb{Z}) = 0 \; \forall i \geq 2\\
	\check{\delta}^{0}(a, b, c) = (b - a, c - a, c - b)\\
	\check{\delta}^{i} = 0 \; \forall i \geq 1 \; .
\end{array}\]
Denoting with $\check{Z}$ the cocycles and with $\check{B}$ the coboundaries, it follows that $\check{Z}^{1}(\mathfrak{U}, \mathbb{Z}) = \check{C}^{1}(\mathfrak{U}, \mathbb{Z})$, while $\check{B}^{1}(\mathfrak{U}, \mathbb{Z}) = \{(n, m, m-n)\}$. Thus, in cohomology of degree 1, $[(n, m, s)] = [(n, m, s) - (n, m, m - n)] = [(0, 0, s - m + n)]$, thus every cohomology class can be represented as $[(0, 0, k)]$. Moreover, $[(0, 0, k)] = [(0, 0, h)]$ if and only if $k = h$. Thus the isomorphism $\check{H}^{1}(X, \mathbb{Z}) \simeq \mathbb{Z}$ can be expressed as $[(0, 0, k)] \simeq k$.

The cohomology class $1$ is thus represented as $[(0, 0, 1)]$, thus, if we consider in $X$ the open cover $f^{-1}\mathfrak{U} = \{f^{-1}U_{0}, f^{-1}U_{1}, f^{-1}U_{2}\}$, the cohomology class $f^{*}1 \in \check{H}^{1}(X, \mathbb{Z})$ is represented by the cocycle with value $1$ on $f^{-1}U_{02}$ and $0$ on $f^{-1}U_{01}$ and $f^{-1}U_{12}$. Let us show that this class corresponds to $c_{1}(f)$. In fact, let us choose the logarithms of $f$ accordingly to the definition of $U_{0}$, $U_{1}$ and $U_{2}$: we define $f\vert_{f^{-1}U_{0}} = e^{2\pi i \rho_{0}}$ with $0 < \rho_{0}(x) < \frac{2}{3}$ for every $x \in U_{0}$, and similarly for $\rho_{1}$ and $\rho_{2}$. We now see that in the definition of $U_{0}$ and $U_{1}$ the angle $\theta$ agrees on $U_{01}$, and the same for $U_{12}$, while there is a difference of $1$ on $U_{02}$. Thus, $\rho_{0} - \rho_{1} = 0$ and $\rho_{1} - \rho_{2} = 0$, while $\rho_{1} - \rho_{2} = 1$. That's why $c_{1}(f)$, computed via $\rho_{i} - \rho_{j}$, agrees with the pull-back of $[(0, 0, 1)]$, which is the class $1$ in $\check{H}^{1}(S^{1}, \mathbb{Z})$. $\square$

\section{Hat-hypercohomology}\label{HatHyperCohomology}

It seems natural to extend the definition of hat-cohomology to the one of hat-hypercohomology, in order to describe connections and holonomies in this way. Actually, we show that the extension can be defined, but it gives no interesting information, therefore, in order to deal with connections, it is better to use Deligne cohomology as we did up to now.

We can apply the general definition of hat-cohomology, given in subsection \ref{Generalization}, to the sheaves of differential forms of a fixed degree on a manifold $X$. What are the cohomology groups $\hat{H}^{*}(X, \Omega^{p}_{\mathbb{R}})$? The case is analogous to the one of the sheaf $\underline{\mathbb{R}} = \Omega^{0}_{\mathbb{R}}$. In fact, since $\Omega^{p}_{\mathbb{R}}$ admits partitions of unity, it is acyclic, i.e.\ $\check{H}^{i}(X, \Omega^{p}_{\mathbb{R}}) = 0 \, \forall i \neq 0$. Thus, in $\check{\rm{C}}$ech cohomology, the only non-trivial group is $\check{H}^{0}(X, \Omega^{p}_{\mathbb{R}})$, which is made by $p$-forms globally defined on $X$. We now show that, in hat-cohomology, $\hat{H}^{0}(X, \Omega^{p}_{\mathbb{R}}) = 0 \, \forall i \in \mathbb{Z}$. In fact, e.g.\ for $i = 0$, it follows from the definition that:
	\[\hat{H}^{0}(X, \Omega^{p}_{\mathbb{R}}) \simeq \check{H}^{0}(X, \Omega^{p}_{\mathbb{R}}) \,/\, \textnormal{homotopy},
\]
where a homotopy between two global forms is a path joining them. Actually, any $p$-form $\omega$ is homotopic to zero, thanks to the following homotopy:
	\[\begin{split}
	\varphi:\; &I \longrightarrow \Lambda^{p}T^{*}X \\
	&\varphi(t)(X_{1}, \ldots, X_{p}) := t \cdot \omega_{x}(X_{1}, \ldots, X_{p}).
\end{split}\]
It is easy to see that $i_{0}^{*} \Omega = 0$ and $i_{1}^{*} \Omega = \omega$, thus $\Omega$ is a homotopy between $\omega$ and $0$. Analogous considerations show that $\hat{H}^{-i}(X, \Omega^{p}_{\mathbb{R}}) = 0$ for any $i \in \mathbb{N}$. In fact, given a cocycle $\varphi:\; I^{i} \rightarrow \Lambda^{p}T^{*}X$, with $\varphi\vert_{J^{i}} = 0$, it holds that $\varphi = \hat{\delta}^{-i-1}\Phi$ for $\Phi$ homotopy between $\varphi$ and $0$ relative to the boundary.

\paragraph{}Let us now study the hypercohomology corresponding to the hat-cohomology. The hat-double-complex associated to the complex of sheaves:
	\[\underline{U}(1) \overset{\tilde{d}}\longrightarrow \Omega^{1}_{\mathbb{R}} \overset{d}\longrightarrow \cdots \overset{d}\longrightarrow \Omega^{p+1}_{\mathbb{R}}
\]
with respect to a good cover $\mathfrak{U}$ is:
\[\xymatrix{
	\cdots \ar[r] & \hat{C}^{-1}(\mathfrak{U}, \Omega^{p+1}_{\mathbb{R}})  \ar[r]^{\hat{\delta}^{-1}} & \hat{C}^{0}(\mathfrak{U}, \Omega^{p+1}_{\mathbb{R}}) \ar[r]^{\hat{\delta}^{0}} & \hat{C}^{1}(\mathfrak{U}, \Omega^{p+1}_{\mathbb{R}}) \ar[r]^(.67){\hat{\delta}^{1}} & \cdots \\
	& \qquad\vdots\qquad \ar[u]^{d} & \qquad\vdots\qquad \ar[u]^{d} & \qquad\vdots\qquad \ar[u]^{d} & \\
	\cdots \ar[r] & \hat{C}^{-1}(\mathfrak{U}, \Omega^{1}_{\mathbb{R}}) \ar[r]^{\hat{\delta}^{-1}} \ar[u]^{d} & \hat{C}^{0}(\mathfrak{U}, \Omega^{1}_{\mathbb{R}}) \ar[r]^{\hat{\delta}^{0}} \ar[u]^{d} & \hat{C}^{1}(\mathfrak{U}, \Omega^{1}_{\mathbb{R}}) \ar[r]^(.67){\hat{\delta}^{1}} \ar[u]^{d} & \cdots \\
	\cdots \ar[r] & \hat{C}^{-1}(\mathfrak{U}, \underline{U}(1)) \ar[r]^{\hat{\delta}^{-1}} \ar[u]^{\tilde{d}} & \hat{C}^{0}(\mathfrak{U}, \underline{U}(1)) \ar[r]^{\hat{\delta}^{0}} \ar[u]^{\tilde{d}} & \hat{C}^{1}(\mathfrak{U}, \underline{U}(1)) \ar[r]^(.67){\hat{\delta}^{1}} \ar[u]^{\tilde{d}} & \cdots
	}
\]
It is infinite on the left and on the right. Let us compute $\hat{H}^{p+1}(X, \underline{U}(1) \rightarrow \Omega^{1}_{\mathbb{R}} \rightarrow \cdots \rightarrow \Omega^{p+1}_{\mathbb{R}})$. We consider for simplicity the case of line bundles, the others are analogous. Thus the complex is:
\[\xymatrix{
	\cdots \ar[r] & \hat{C}^{-1}(\mathfrak{U}, \Omega^{1}_{\mathbb{R}}) \ar[r]^{\hat{\delta}^{-1}} & \hat{C}^{0}(\mathfrak{U}, \Omega^{1}_{\mathbb{R}}) \ar[r]^{\hat{\delta}^{0}} & \hat{C}^{1}(\mathfrak{U}, \Omega^{1}_{\mathbb{R}}) \ar[r]^{\hat{\delta}^{1}} & \cdots \\
	\cdots \ar[r] & \hat{C}^{-1}(\mathfrak{U}, \underline{U}(1)) \ar[r]^{\hat{\delta}^{-1}} \ar[u]^{\tilde{d}} & \hat{C}^{0}(\mathfrak{U}, \underline{U}(1)) \ar[r]^{\hat{\delta}^{0}} \ar[u]^{\tilde{d}} & \hat{C}^{1}(\mathfrak{U}, \underline{U}(1)) \ar[r]^{\hat{\delta}^{1}} \ar[u]^{\tilde{d}} & \cdots
	}
\]
and the 1-cochains are the same as for $\check{\rm{C}}$ech hypercohomlogy, i.e.\ $\hat{C}^{1}(\mathfrak{U}, \underline{U}(1)) \oplus \hat{C}^{0}(\mathfrak{U}, \Omega^{1}_{\mathbb{R}})$, which is equal by definition to $\check{C}^{1}(\mathfrak{U}, \underline{U}(1)) \oplus \check{C}^{0}(\mathfrak{U}, \Omega^{1}_{\mathbb{R}})$. Since also the coboundaries are equal for non-negative degrees, the cocycles are the same: they are of the form $\{g_{\alpha\beta}, - A_{\alpha}\}$ with $g_{\alpha\beta}g_{\beta\gamma}g_{\gamma\alpha} = 1$ and $A_{\alpha} - A_{\beta} = \frac{1}{2\pi i} d \log g_{\alpha\beta}$. The difference can be seen in the coboundaries: there are the usual coboundaries coming from the cochains in $\hat{C}^{0}(\mathfrak{U}, \underline{U}(1))$, but there are also the coboundaries coming from cochains in $\hat{C}^{-1}(\mathfrak{U}, \Omega^{1}_{\mathbb{R}})$. Thus the general coboundary is of the form $\delta(\{g_{\alpha}\}, \varphi) = (\{g_{\alpha}^{-1}g_{\beta}, \frac{1}{2\pi i} d \log g_{\alpha} + \varphi(1)\}$, for $\varphi: I \rightarrow T^{*}X$. This means that there are new coboundaries of the form $(1, A)$ for $A$ a globally-defined $1$-form: but these are exactly the topologically trivial line bundles, with any connection. Thus, the cohomology class contains information only about the topology of the line bundle, i.e.\ there is an isomorphism $\hat{H}^{1}(X, \underline{U}(1) \rightarrow \Omega^{1}_{\mathbb{R}}) \simeq \hat{H}^{1}(X, \underline{U}(1))$. For higher degrees the situation is analogous. In degree $-1$ it is even clearer since the double-complex has only one line and it reduces to the hat-complex, thus we get exactly $\hat{H}^{0}(X, \underline{U}(1))$. Thus, in general:
	\[\hat{H}^{p+1}(X, \underline{U}(1) \rightarrow \Omega^{1}_{\mathbb{R}} \rightarrow \cdots \rightarrow \Omega^{p+1}_{\mathbb{R}}) \;\simeq\; \hat{H}^{p+1}(X, \underline{U}(1)) \quad \forall p \geq -1 \; .
\]
For $i \leq -2$ the double complex is the $0$-complex, thus, for $i \leq -3$, the isomorphism still holds, but not for $i = -2$, since $\hat{H}^{-1}(X, \underline{U}(1)) \simeq \mathbb{Z}$. Actually, one can also consider the infinite complex of sheaves:
	\[\underline{U}(1) \overset{\tilde{d}}\longrightarrow \Omega^{1}_{\mathbb{R}} \overset{d}\longrightarrow \Omega^{2}_{\mathbb{R}} \overset{d}\longrightarrow \cdots
\]
and prove in the same way that:
	\[\hat{H}^{p+1}(X, \underline{U}(1) \rightarrow \Omega^{1}_{\mathbb{R}} \rightarrow \Omega^{2}_{\mathbb{R}} \rightarrow \cdots) \;\simeq\; \hat{H}^{p+1}(X, \underline{U}(1)) \quad \forall p \in \mathbb{Z} \; .
\]
Now the isomorphism holds also for $p = -2$.


\end{document}